%% file: MAIN.tex
\documentclass[journal=nalefd,manuscript=letter,layout=twocolumn]{achemso} 
\usepackage{amsmath}
\usepackage{amssymb}
\usepackage{graphicx}
\usepackage{esint}
\usepackage{siunitx}
\usepackage{color}

\setkeys{acs}{doi = true}

\definecolor{olive}{rgb}{0,0.8,0}

\usepackage{ulem} 

\title{Luminescence properties of closely packed organic color centers grafted on a carbon nanotube}

\author{Antoine Borel}
\author{Federico Rapisarda}
\affiliation{Laboratoire de physique de l'ENS, Université PSL, CNRS, Sorbonne Université, Université Paris Cité,75005 Paris, France}

\author{Stephen K. Doorn}
\affiliation{Center for Integrated Nanotechnologies, Materials Physics and Applications Division, Los Alamos National Laboratory, Los Alamos, New Mexico 87545, United States}

\author{Christophe Voisin}
\author{Yannick Chassagneux}
\email{yannick.chassagneux@phys.ens.fr}
\affiliation{Laboratoire de physique de l'ENS, Université PSL, CNRS, Sorbonne Université, Université Paris Cité,75005 Paris, France}

\keywords{Carbon nanotube, Organic Color Center, Spectral Diffusion, Coupled Emitters}

\begin{document}


\begin{abstract}
We report on the photo-luminescence of pairs of organic color centers in single-wall carbon nanotubes grafted with 3,5 dichlorobenzene. Using various techniques such as intensity correlations, super-localization microscopy or luminescence excitation spectroscopy, we distinguish two pairs  of color centers grafted on the same nanotube; the distance between the pairs is on the order of several hundreds of nanometers. In contrast, by studying the strong temporal correlations in the spectral diffusion in the framework of photo-induced Stark effect, we can estimate the distance within each pair to be of the order of a few nanometers. Finally, the electronic population dynamics is investigated using time-resolved luminescence and saturation measurements, showing a biexponential decay with a fast overall recombination (compatible with a fast population transfer between the color centers within a pair) and a weak delayed repopulation of the traps possibly due to the diffusion of excitons along the tube axis. 
\end{abstract}

Color centers stand for pseudo atomic systems embedded in a solid-state material that can be manipulated optically for coherent control, single photon emission and other quantum operations \cite{dolan_robust_2018,wang_coherent_2020,dolde_room-temperature_2013}. They stand up as valuable building blocks for quantum technologies, with all the advantages associated to solid-state systems. Among many possible approaches, the case of quantum emitters in close vicinity ($\delta \sim \text{nm}$) is of particular interest for sensing \cite{Ji2024, Tom24} or to implement quantum superpositions and entangled states yielding for instance super and sub-radiance effects... Such effects have been demonstrated using incidental pairs of quantum dots \cite{koong_coherence_2022} or pairs of molecules  \cite{hettich_nanometer_2002,trebbia_tailoring_2022}. In this respect, the control of the relative position of the emitters at the nanoscale would open the way to a better control over the fabrication of coupled quantum emitters, for applications in solid-state quantum technologies. 

Carbon nanotubes (CNTs) have recently been considered for such applications due to the possibility to use them as a functionalizable backbone to graft various molecules using organic chemistry tools such as diazonium chemistry \cite{Piao2013,he_tunable_2017,ma_room-temperature_2015,berger_brightening_2019, settele_synthetic_2021}. The molecule grafted on a semi-conducting  nanotube acts as a deep potential well ($\sim 100$~meV) for the excitons generated across the bandgap, yielding the concept of organic color centers (OCC) which proved to be an original single photon source in the telecom bands up to room temperature \cite{he_tunable_2017}. This 1D backbone increases drastically the probability to generate pairs of closely packed emitters (much closer than the wavelength $\lambda$) since it   scales like $\delta/\lambda$ in 1D versus   $(\delta/\lambda)^3$ in a    3D case, yielding a $10^4$ fold enhancement. This concept was recently explored for guanine decorated CNTs, where closely spaced OCCs could be observed, giving rise to anti-correlation signatures \cite{zheng_quantum_2021}.

In this work, we report on the emblematic observation of two couples of OCC along the same nanotube, showing strong spectral correlations, which we analyse in the framework of Stark induced spectral diffusion. We obtain an estimate of the distance between OCCs within each pair of the order of a few nanometers. Using saturation and time-resolved spectroscopy, we investigate the population dynamics  within each pair of OCC. 

\textbf{Methods.}
The sample consists of polymer (PFO) wrapped (7,5) nanotubes functionalized with 3,5 dichlorobenzene diazonium salt by the dip doping method \cite{he_tunable_2017}. The 3,5 dichlorobenzene ion is covalently attached to the CNT surface by converting a $sp^2$ into a $sp^3$ bound. A 160 nm thick layer of polystyrene is then deposited over the nanotubes.   In this study, we focused  on the subclass $E_{11}^{*-}$ of the photoluminescent states of the OCC, that emit in the O band \cite{he_tunable_2017}.

The experimental setup consists in a home-made microscope and the sample is cooled at 9K \cite{jeantet_widely_2016,Borel2023}. After being excited with a 800 nm laser, the photo-luminescence (PL) was analyzed either with a spectrometer or with a fibered  single photon detector (SSPD). The CNT is coupled to a fiber Fabry-Perot cavity, as described in \cite{Borel2023}. Here, the cavity is used primarily as a sharp tunable band-pass filter. 

\begin{figure}[ht!]
\includegraphics[width=3.33in]{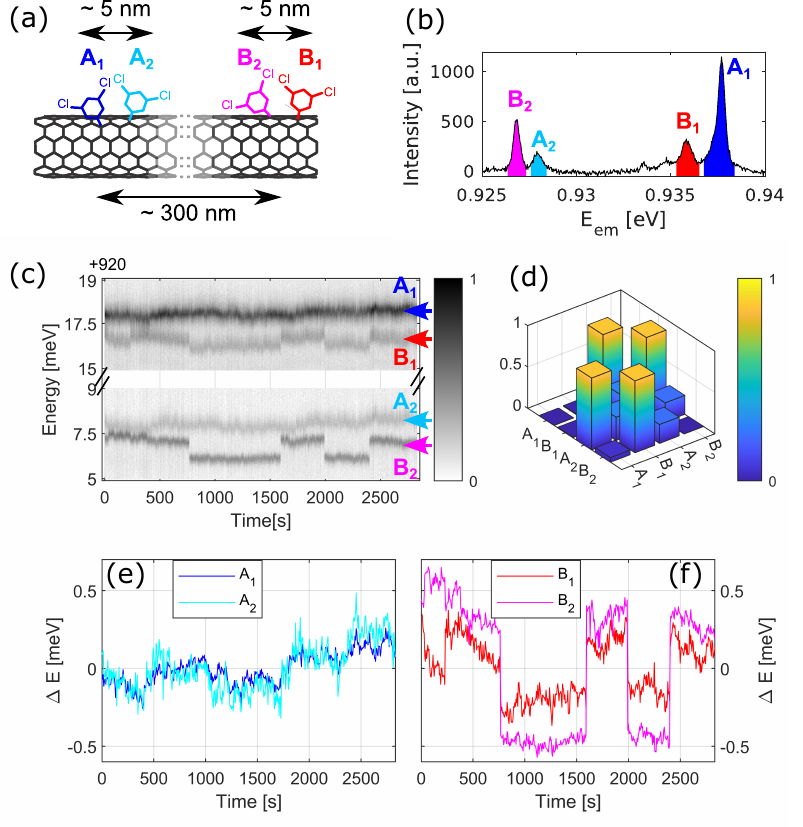} \caption{ (a) Sketch of the system (b) Typical luminescence spectrum (excitation wavelength 800~nm, power density 4.2 kW/cm$^2$  
showing four main lines within a 20~meV window. (c) Time trace of the PL spectrum with a 1 s time bin showing correlated spectral wandering and spectral jumps. (d) correlation matrix of the fluctuating emission energies of the four lines as labeled in (b)  computed from temporal traces (c). (e) and (f) time trace of the energy shifts for the emitters in pair A and pair B respectively.   \label{fig:Presentation}}
\end{figure}

\textbf{Observations.}
We investigated a 20~meV band around 0.93~eV. Beside numerous spectra consisting in a single line, we happened to observe more complex spectra such as that of  Fig.~\ref{fig:Presentation}(b), which consists of 4 lines.  The evolution of the spectrum over time  is shown in Fig.\ref{fig:Presentation}~(c). Correlated spectral wandering and jumps can be observed.  To assess the correlations between these lines, their spectral position correlation matrix is shown Fig.\ref{fig:Presentation}~(d). The emission energies are temporally correlated for lines $A_1$ and $A_2$ (Fig.\ref{fig:Presentation}~(e)) in the one hand and lines $B_1$ and $B_2$ on the other hand (Fig.\ref{fig:Presentation}~(f)).  
 
Such correlations may have several origins. They are either the signature of two very close emitters sharing the same electrostatic environment and thus having correlated Stark shifts, or the signature of a single emitter showing two bright levels or a single emitter subject to fast switching between two subsystems due to the interaction with a nearby fast fluctuator.  In the following,  we show that the spectrum of Fig.~\ref{fig:Presentation}(b) actually arises from two pairs of closely packed emitters as depicted in Fig.~\ref{fig:Presentation}(a). 

\begin{figure}[ht!]
    \centering
    \includegraphics[width=3.33in]{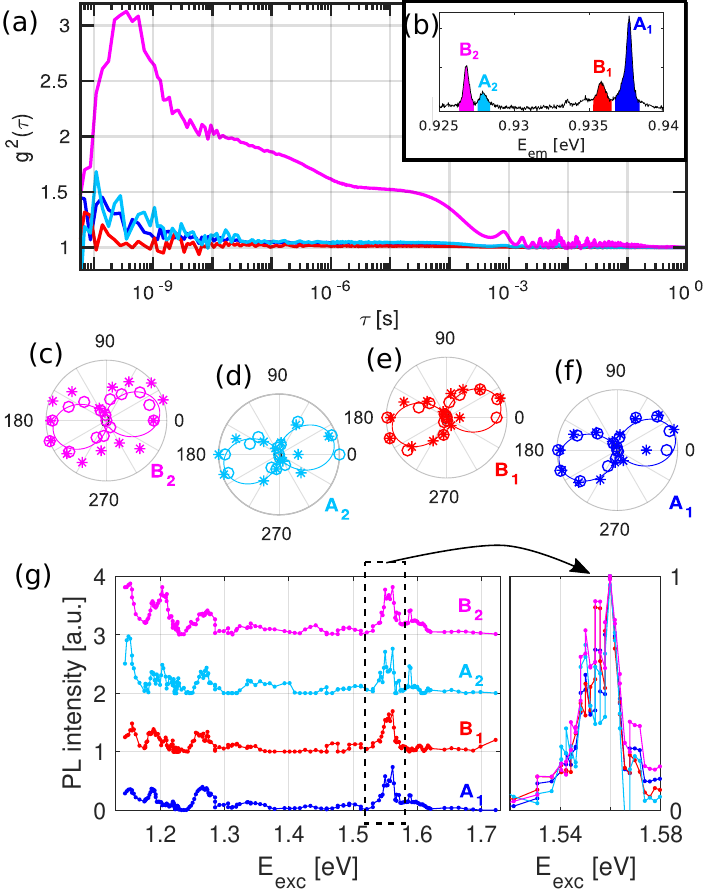}
    \caption{ (a) Intensity auto-correlation function under cw excitation of each lines according to the color code of panel (b). (c-f) Polarization diagrams of each PL line (open symbols for the PL signal, full symbols for the excitation) showing the copolarization of the four emitters. (g) Excitation spectrum of each of the 4 lines according to the color code in (b) showing common resonances at 1.56 eV (close-up in the inset) and 1.265 eV. }
    \label{fig:g2polaPLE}
\end{figure}

Eventhough the dark/bright splitting reported for both pristine and aryl-functionalized (7,5) CNT (4 meV to 15 meV) under high magnetic field ($>$ 5 T) \cite{gandil_spectroscopic_2019, kim_hidden_2020} roughly matches the splittings we observe for pair $A$ and pair $B$ (Fig.~\ref{fig:Presentation}(b)), we rule out the interpretation because in the absence of a magnetic field the intensity contrast would be much larger than observed here. In addition, the observation of such split lines in our sample is very unusual, at contrast with an intrinsic dark/bright splitting which should be ubiquitous. A fluctuator could split the excitonic emission energy on a time scale much shorter than the acquisition time. The fluctuations of these lines on the longer (1 s) time-scale would naturally be correlated since both lines arise from the same object. The intensity auto-correlation function should reflect the switching timescale. The autocorrelations of both lines should exhibit a step at the same time delay (an exponential decay appears as a step  in semilogx  scale); the product of the two step amplitudes being equal to one (see model in SI). Fig.~\ref{fig:g2polaPLE}(a) shows the auto-correlations of each of the four lines. Although steps can be identified, their amplitudes do not fulfill the condition described above, ruling out the hypothesis of a fast fluctuator. In addition, for such fast switching scenario, one would also observe cross-correlations in the intensity fluctuations or cross-correlations between intensities and energies \cite{boiron_are_1999}. This is not the case here, as shown by the correlation matrices (SI).  Finally, we show below that each line has its own decay dynamics, which tends to confirm that they arise from physically different OCCs.  Hence, we rule out the scenario of a fast switching system and we are left with the simplest explanation: each pair of lines arises from two distinct OCCs that are close enough to probe the same fluctuating electrostatic environment.   

The diffraction limited resolution is not sufficient to decide whether the four OCCs actually belong to the same nanotube or to  distinct ones (within the diffraction spot). To investigate this question further, we first checked their PL polarization diagrams. In fact, both their intrinsic PL and that arising from nearby emitters \cite{roquelet_local_2012} are polarized along the nanotube axis.  The diagrams are shown in Fig.~\ref{fig:g2polaPLE}~(c-f), together with a dipolar diagram fit. The four PL spectral lines are co-polarized with the same angle $14^\circ \pm 4^\circ$. The
same copolarization signature is observed on the excitation diagrams (full symbols in Fig.~\ref{fig:g2polaPLE}(c-f)). This strongly suggests that the four OCCs are attached to the same nanotube and are fed by the excitons generated at higher energy in the nanotube. 

To confirm this interpretation, we performed an excitation resolved PL study (PLE) displayed in Fig.~\ref{fig:g2polaPLE}(g). The four emitters show the same resonances especially at about 1.56 eV, in line with a common feeding process. Finally, we performed a spectrally resolved super-localization measurement similarly to \cite{raynaud_super-localization_2019} in order to possibly determine a sub-wavelength localization of the emission sites. The spatial precision of such measurement  was typically 15 nm to 25 nm for optimized setups such as in \cite{raynaud_super-localization_2019} or \cite{Danne18}. Here we could only resolve the brightest emitters $A_1$ and $B_2$ and we measured a distance of 300 nm $\pm$ 150 nm between them (see SI).  The distance within the pairs could not be resolved with this technique.  

\textbf{Correlated Stark shifts.}
We used the correlated Stark shifts to evaluate the typical distance between the two emitters.  Within the framework of the linear Stark effect, the energy shifts are given by: $\Delta E = -\Vec{p}\cdot \Vec{E}$, where $\Vec{E}$ is the electric field and $\Vec{p}$ a permanent dipole associated with the OCC. Glückert et al.  \cite{gluckert_dipolar_2018} showed that an oxygen OCC has a permanent dipolar moment of the order of 0.3 e.\AA{} (where $e$ is the elementary charge). This permanent dipole has to be differentiated from the transition dipole. In particular, it is not necessarily oriented along the CNT axis but would rather point along the direction of the molecular substituent making up the OCC (we note the exact orientation of the dipole is not a critical factor in our modelling).  As a first approach, we computed the Stark shifts using the permanent dipole moment of 3,5 dichlorobenzene (0.47 e.\AA{}).  

This dipole is in interaction with the electrostatic environment. Using the superposition principle, the trapping/detrapping of an elementary charge in the vicinity of the dipole yields a Stark shift : 
\begin{equation}
    \delta E_{stark}=\pm\frac{e}{4\pi \epsilon_0 \epsilon_r} \frac{\Vec{p}\cdot\Vec{r}_{p, e}}{r^3_{p, e}}, \label{eq:stark}
\end{equation}
where $ \epsilon_{r}$ is the polystyrene static dielectric permittivity, $\Vec{p}$ is the permanent dipole moment, $\vec{r}_{p,e}$ is the vector between the positions of the dipole and the charge, the positive (resp. negative) sign corresponds to a trapping (resp. detrapping) event. Experimentally, $\delta E$ represents the spectral shift between two consecutive acquisitions. We observe sharp spectral jumps compatible with individual charge trapping events.
 
\begin{figure}[ht!]
    \includegraphics[width=3.33in]{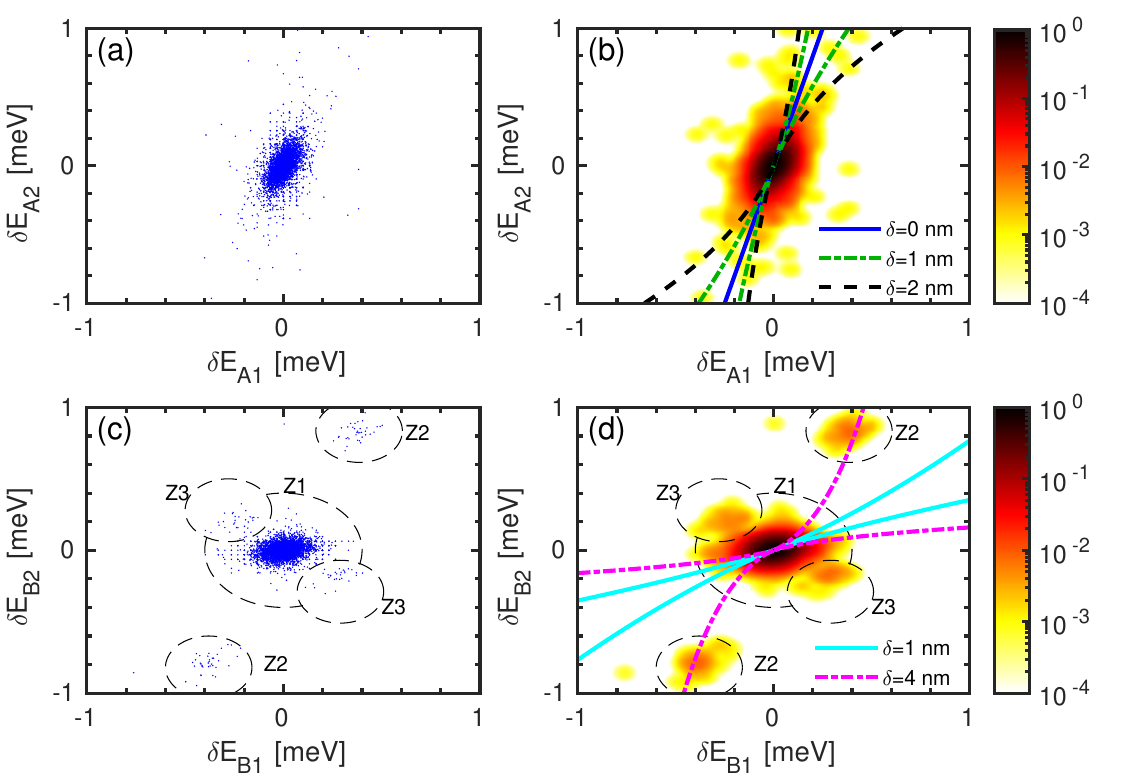} \caption{  (a) Spectral shifts for the emitters of pair A : each point corresponds to the energy shift between consecutive spectra of line $A_2$ function of the shift of line $A_1$ (4672 points).  (b) Log-scale density plot generated from panel (a) (see text). The blue, green and black lines correspond to  the model (eq ~\ref{eq:stark}), assuming that $\vec{p}_{A1}.\vec{e}_z=0.25|p_0|$; $\vec{p}_{A2}.\vec{e}_z=|p_0|$ , $\vec{e}_z$ being the unit vector along the nanotube axis. We use $p_0=0.47 e\text{\AA}$ and $\epsilon_r=2.25$. The dipole separation  is  $\delta=0,1 \text{ and }2$ nm. (c) and (d) Similar to (a) and (b) for the emitters of pair B. Five distinct clouds of points (symmetrical with respect to the origin) are distinguishable hereafter called lobes Z1, Z2 and Z3. The cyan and magenta lines correspond to dipole separation $\delta=1 \text{ and } 4$ nm with $\vec{p}_{B1}.\vec{e}_z=|p_0|$ and $\vec{p}_{B2}.\vec{e}_z=0.5|p_0|$  indicating a finite separation of the emitters.}\label{fig:densiteelectrostatique}
    \end{figure}

In addition, we observed that the magnitude of the spectral diffusion increases with the optical power, pointing towards photo-generated charges. Since the polystyrene matrix is transparent, absorption mainly occurs in the  nanotube, with charges being possibly transferred to the close vicinity  due to photo-ionisation. Hence, in a first approach, we chose a quasi 1D model where the fluctuating charges are bound to the nanotube axis. For each line, the energy difference between consecutive spectra are represented in a bi-dimensional plot (Fig.~\ref{fig:densiteelectrostatique}(a,c)). To account for the  resolution of our spectrometer, we represented each point as a 2D Gaussian spot with a standard deviation of 33~µeV. After summing up all these contributions, we obtain a density plot representing the dual probability of spectral jumps of the two lines of each pair (Fig.~\ref{fig:densiteelectrostatique}(b,d)). The two pairs show contrasted distributions. For the pair A, the energy shifts $\delta E_{A1}$ and $\delta E_{A2}$ are linearly correlated. For pair B, the density plot shows five main lobes with a symmetry center in (0,0). The pair of lobes that are symmetrical with respect to the origin correspond to finite spectral shifts of either identical signs (lobes Z2) or opposite signs (lobes Z3). We modeled this distribution assuming that each point corresponds to the trapping or de-trapping of an elementary charge in the vicinity of the nanotube resulting in a shift given by eq.\ref{eq:stark}.   Small energy shifts are related to remote charges and the deviation of the slope to one of the central lobe reflects the different projections of the permanent dipoles along the nanotube axis.
For pair A, this analysis imposes an upper bound (obtained when one dipole is aligned with the nanotube) on the distance between the two dipoles $\delta \lesssim 2$~nm. 
In fact, for larger dipole separation a splitting of the central lobe is expected, due to an increasing differential effect of a charge on the left and right side OCC. In addition, the absence of side lobes also points to a low distance between the OCCs. In particular, the absence of the anti-symmetric lobes rules out a configuration with a charge trapped in-between the OCCs and thus points to a tiny separation within pair A.

For pair B, we can formally exclude that the two dipoles are located at the same position since correlated shifts (lobes Z2) and anti-correlated shifts (lobes Z3) are observed, which corresponds to the case where the trapped charge is located between the two OCC. Lobe Z1 results from fluctuating  charges located far away from the pair. Lobe Z2 corresponds to a peculiar nearby trapping site on the CNT axis. We can deduce the dipole-dipole separation in pair B to be of the order of 4 nm. In a 1D model, lobe Z3 would correspond to a single trapping site between the two dipoles. However, in this configuration, the distances between the charge and the dipoles is comparable to the diameter of the nanotube and the 1D approximation has to be reconsidered.  
By considering the 3D geometry of the nanotube and different dipole orientations, the spectral shifts observed in Fig.\ref{fig:densiteelectrostatique}~(c,d) can be reproduced (SI).

We end up with the picture of two pairs of OCCs separated by $\sim 300\ \text{nm}$ whereas the distance within each pair is on the order of a few nanometers. This tiny separation between the OCCs in each pair is likely to impact the dynamics of the system \cite{zheng_quantum_2021, zheng_photoluminescence_2021}, which we investigate in the following section.

\begin{figure}[ht!]
    \includegraphics[width=3.33in]{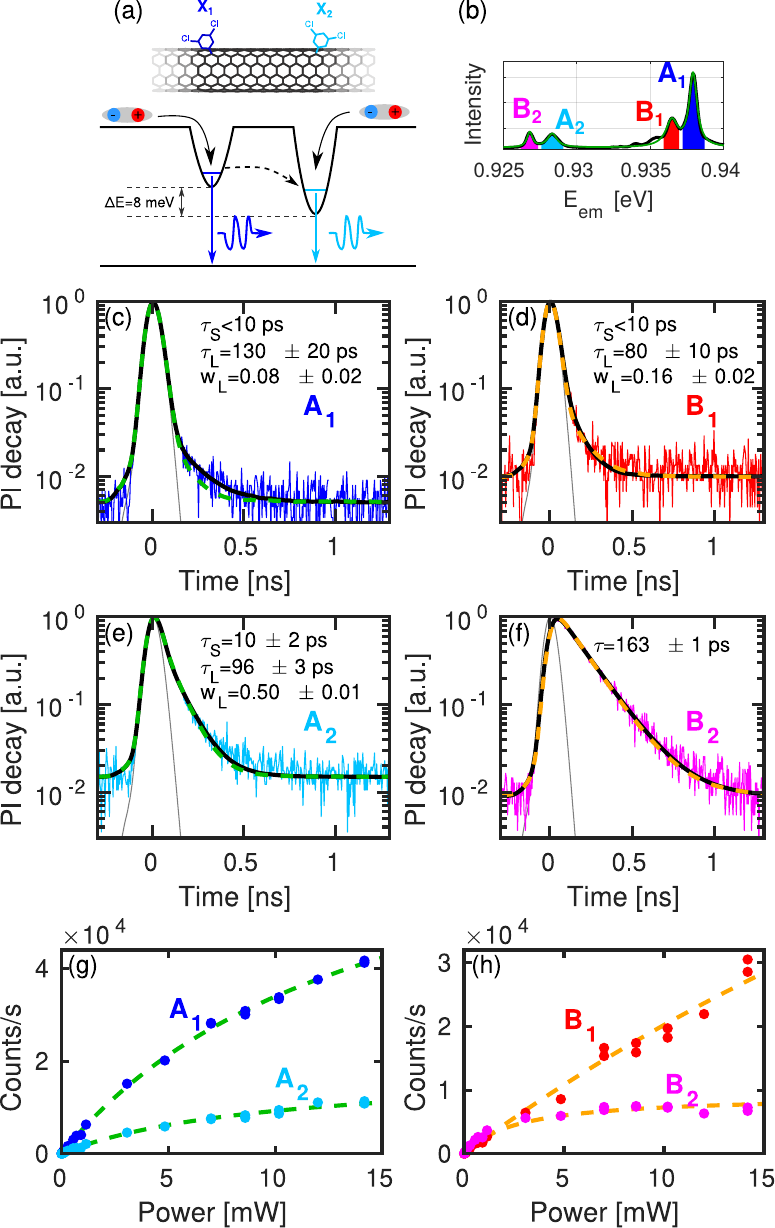} \caption{(a) Sketch of a pair of OCCs showing possible tunnel coupling. (b) PL spectrum showing the color code for labelling the lines. (c-f) Semi-log plot of the PL decay dynamics for the spectrally filtered lines, according to the color code of (b) The black line corresponds to a bi-exponetial fits (mono-exponential for (f) ). The lifetimes  $\tau_L$ and $\tau_S$ were extracted from the fits and  $\text{w}_\text{L}$ indicates the integrated weight of the long component. The instrument response function is indicated by the grey line. (g) (resp. (h)) Saturation measurements of each line of pair A (resp. B). For pair A (resp. B), the green (resp. orange) dashed lines of panel (c), (e) and (g) (resp.  (d),(f) and (h)) correspond  to a simultaneous fitting of the two decays and of the two saturation measurements for each pair according to the four level model presented in the main text.  \label{fig:Dynamique}}
    \end{figure}

\textbf{Population dynamics.}
To measure the PL decay, we exploit the coupling of the nanotube to a tunable fiber Fabry-Perot cavity, as described in \cite{Borel2023}, allowing us to obtain a good spectral rejection with a typical width of 50 µeV. The PL transients are shown in Fig.~\ref{fig:Dynamique}(c-f) together with bi-exponential fits.  For both pairs of OCCs, the upper energy line shows the fastest weighted average decay. For the pair $A$, the decay times of each emitter are similar, yet the weight of the longer decay is 6 times greater for the lower energy level. According to literature the lifetime of (7,5) aryl-functionalized CNT at 10K is typically between 100-200 ps with a dominating contribution of the longer component (90\%)  \cite{he_tunable_2017,kim_photoluminescence_2019}. Emitters $A_1$ and $B_1$ (upper levels in each pair) thus exhibit unusually fast decay revealing possible interaction with other levels. However, even line $A_2$ is significantly faster than previously reported, especially regarding the significant weight of the short decay component. In fact,  $A_1, A_2 \text{ and } B_1$ show similar behavior with a dominant fast decays (below our resolution of 10 ps) and a small weight  $\sim 100$ ps long component.  For each pair of OCCs, we model this behavior by a four level scheme with a ground state, an upper reservoir level pumped by the laser that populates the two OCCs levels (Fig.~\ref{fig:Dynamique} (a)). Each level has its own decay time and we include a possible population transfer between the levels. This coupling simply results in reducing the lifetime of the upper level, in agreement with the observations. However, it cannot explain the fast component of the lower level as it rather results in a finite rise time, which we cannot observe due to our limited time-resolution.  The high energy excitation scheme  is assumed to populate both the OCC levels and a  reservoir repopulating the OCC levels over a 100~ps time scale. This reservoir can be either the $E_{11}$ excitonic level of the nanotubes (in agreement with $E_{11}$ lifetime in the literature \cite{jeantet_widely_2016})   , or OCC dark states  \cite{kim_photoluminescence_2019} , giving rise to the long decay observed in all transient. The atypical value measured for the short decay component, which basically corresponds to the OCC lifetime for $A_1$, $A_2$ and $B_1$, may be caused by an unusually fast  non-radiative decay for this specific OCC. In contrast, $B_2$ can be accurately modeled by a mono-exponential decay of the order of 150 ps. Here, it is important to note that if the OCC lifetime is longer than the repopulation time , the overall decay is very similar to a mono-exponential decay.  This also explains why $g^{(2)}(0)$ can be very low for long-lived OCCs\cite{he_tunable_2017}.
To investigate the system dynamics from a different perspective, we performed saturation measurements under cw excitation as shown in Fig.~\ref{fig:Dynamique}~(g,h). The saturation curves are well reproduced by the usual behavior $I\propto P/(P+P_\text{sat})$, where the saturation power are  $P_{sat,A_1}=14 \pm 2 \text{ mW}$, $P_{sat,A_2} = 11 \pm 2 \text{ mW}$ and $P_{sat,B_2} = 1.6 \pm 0.3 \text{ mW}$, whereas $P_{sat,B_1}$ could not be extracted for this range of excitation densities. Within a pair, the ratio of the saturation power should reflect the ratio of the decay rates of the corresponding levels. For pair $B$, the ratio of $P_{\text{sat}}$ is at least larger than 15, thus confirming the hypothesis that the lifetime of $B_1$ corresponds mainly to the short component in the biexponential decay. In contrast emitters $A_1$ and $A_2$ show relatively similar saturation thresholds, indicating that their unresolved rapid decay component are comparable.  

We could reproduce simultaneously both the decay and the saturation curves of each pair using the four-level scheme (Fig.~\ref{fig:Dynamique} (a)) as shown by the green (resp. orange) dashed lines for pair $A$ (resp. $B$) in Fig.~\ref{fig:Dynamique}~(c-h). We attempted to determine if the fast decay of the upper levels could be related to an efficient tunnel coupling of the population to the lower level in each pair (especially considering the small intra-pair OCC distance of pair A). This scenario is compatible with the data. However, our limited time-resolution prevents us from observing directly the rise time of the lower level signal that is expected in this scenario.


In conclusion, we presented a thorough investigation of an original system composed of 4 OCCs, the emission energies of which are temporally correlated by pairs. Long time-scale intensity auto-correlations show that the system cannot be described with fewer emitters hosting several states. We showed that these emitters are most certainly all attached to the same CNT. Using a thorough analysis and modeling of the correlated Stark shifts, we could distinguish two closely packed pairs separated by several hundreds of nm, while the distance between OCCs within a pair is on the order of a few nm.  Finally, the dynamics and saturation  show rather untypical behaviors for the upper levels with unusually fast decays, however the resolution is not sufficient to prove unambiguously that this is due to a population transfer to the lower levels; this hypothesis remains however plausible, in view of the inter-OCC distances within a pair, which are comparable to the exciton Bohr radius in nanotubes. This study reveals that grafted OCCs on the backbone of a CNTs is a promising way to build ensembles of closely packed quantum emitters, with possible electronic coupling. The strange coincidence of finding two such pairs within a diffraction spot, raises the question whether the grafting would be a collaborative process, possibly fostered by the lattice deformation induced by the first OCC. The question also comes if inter-level coupling could be achieved (at shorter distance), paving the way to advanced quantum manipulations of coupled OCC states.

\subsubsection*{Funding sources}
This work was supported by the French National Research Agency (ANR) through the grant DELICACY (ANR-22-CE47-0001), HETEROBNC (ANR-20-CE09-0014) and by the FLAG-ERA Grant OPERA (DFG 437130745 and ANR-19-GRF1-0002).

\begin{acknowledgement}
The authors thanks  X. He for sample preparation and C. Diederichs, E. Baudin, G. Hetet, Z. Said for fruitful discussion. 
\end{acknowledgement}
%


\input{MAIN.bbl}

\end{document}


\title{Supporting Information:  Luminescence properties of closely packed organic color centers grafted on a carbon nanotube}

\author{Antoine Borel}
\author{Federico Rapisarda}
\affiliation{Laboratoire de physique de l'ENS, Université PSL, CNRS, Sorbonne Université, Université Paris Cité, Paris, France}

\author{Stephen K. Doorn}
\affiliation{Center for Integrated Nanotechnologies, Materials Physics and Applications Division, Los Alamos National Laboratory, Los Alamos, New Mexico 87545, United States}

\author{Christophe Voisin}
\author{Yannick Chassagneux}
\email{yannick.chassagneux@phys.ens.fr}
\affiliation{Laboratoire de physique de l'ENS, Université PSL, CNRS, Sorbonne Université, Université Paris Cité, Paris, France}


\maketitle

\renewcommand{\thefigure}{S-\arabic{figure}}

%

\section{Temporal correlations} 


\subsection{Grafted SWCNT}

We recorded a series of several thousand spectra each with an integration time of 1s (time traces) for several excitation powers. We fitted each emission line by a Lorentzian function and extracted their emission energies and intensities. We computed the Pearson correlation coefficient between these parameters. For the sake of readability, we artificially set the auto-correlation coefficients to 0. The spectral fluctuations increase with excitation power density, pointing to a photo-induced effect, possibly mediated by photo-generated charges.

\begin{figure}[!ht]
    \centering

    \includegraphics[scale=0.5]{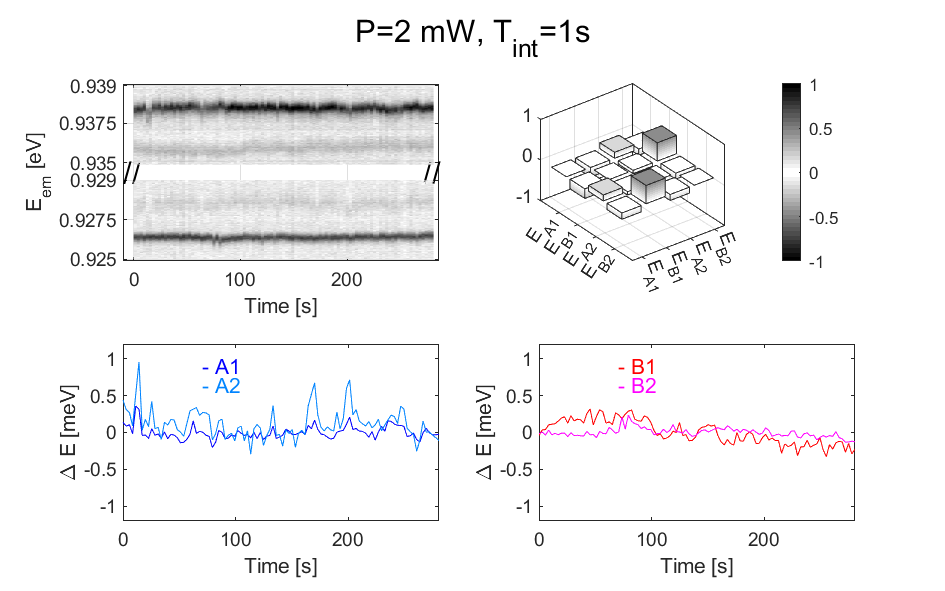}
    \includegraphics[scale=0.5]{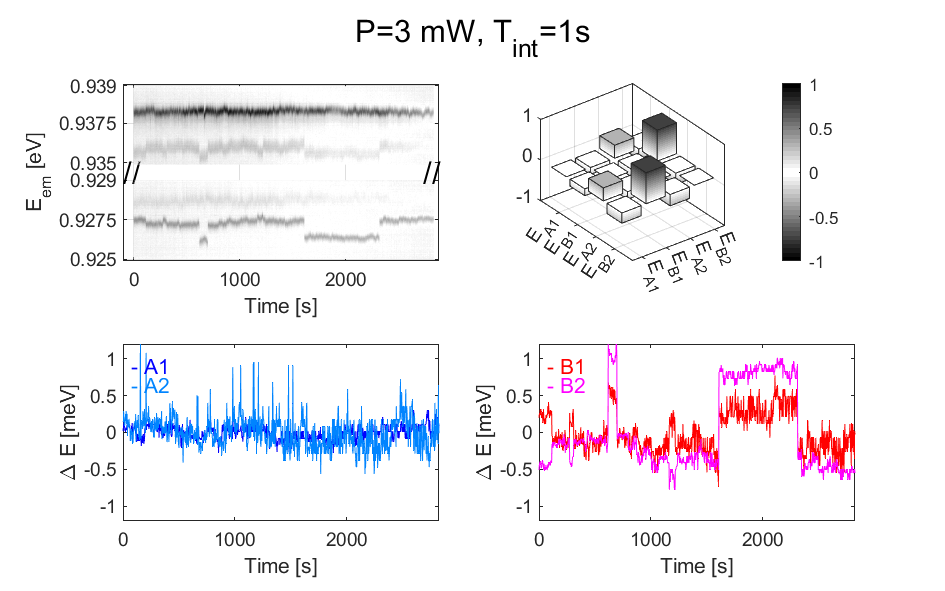}
    \includegraphics[scale=0.5]{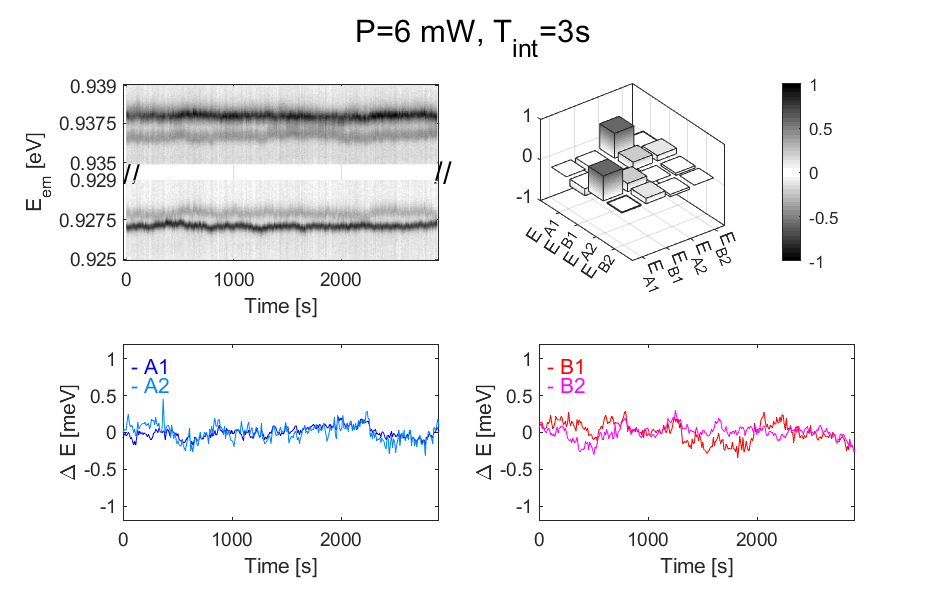}
    \includegraphics[scale=0.5]{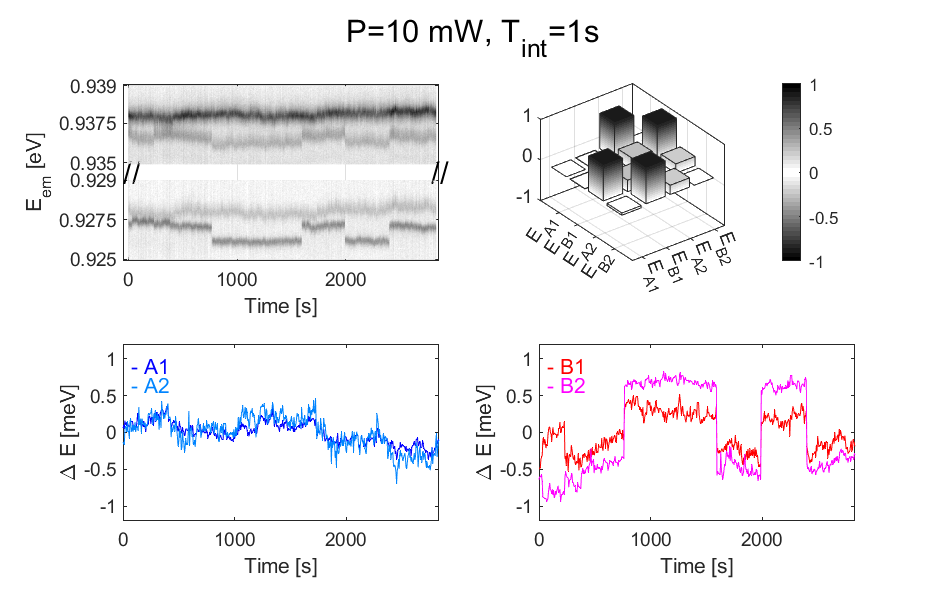}
    \includegraphics[scale=0.5]{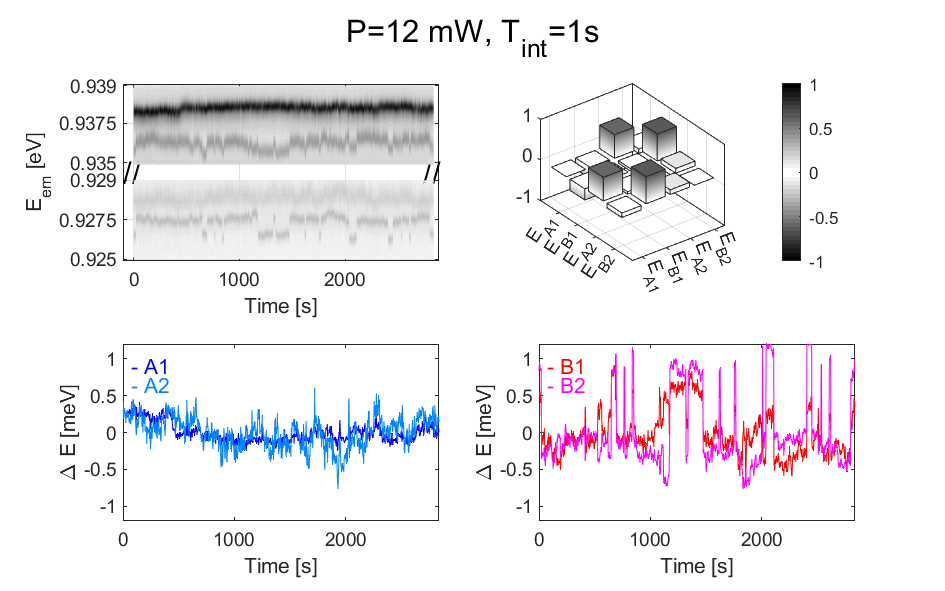}
    \includegraphics[scale=0.5]{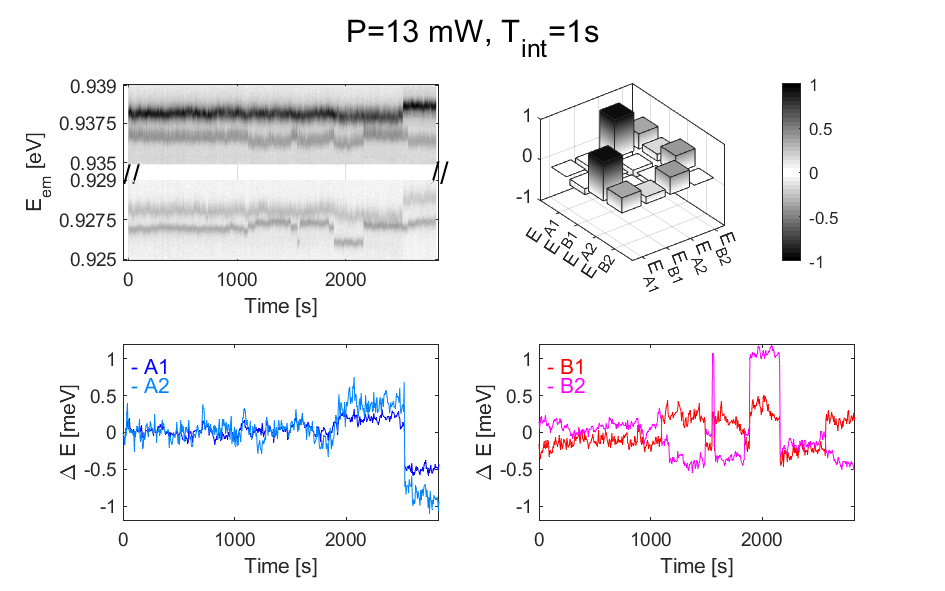}
    
    \caption{Temporal correlations between emission energies for pair A and B for different excitation power. For each power, the time trace is displayed as a map in gray color on a reduced scale. Energy deviation to the mean value $\Delta E(t)=E(t)-<E>$ is then displayed for both pair (coloured 2D plot) before computing the Pearson correlation coefficients.}
    \label{fig:Ecorrelations}
\end{figure}

\begin{figure}
    \centering

    \includegraphics[scale=0.5]{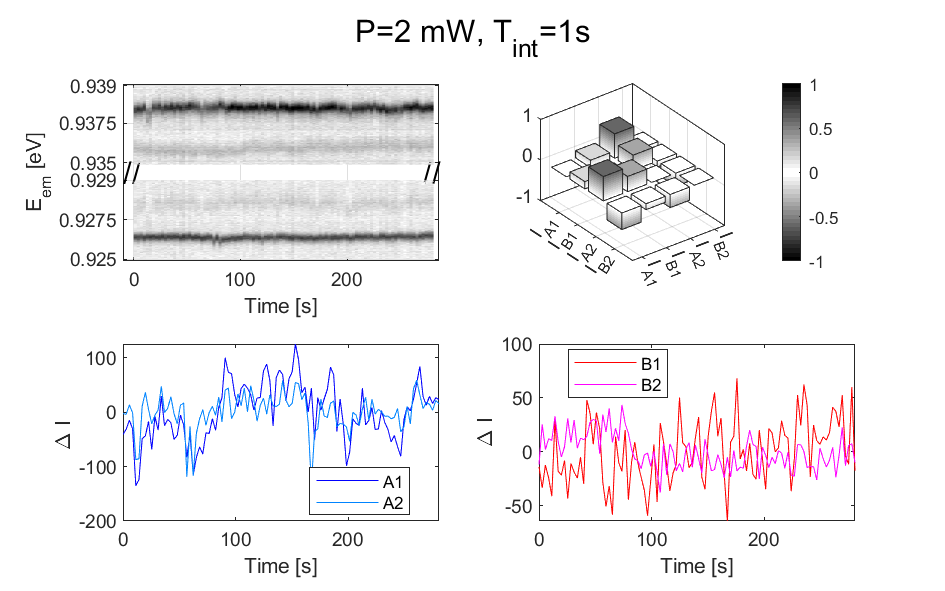}
    \includegraphics[scale=0.34]{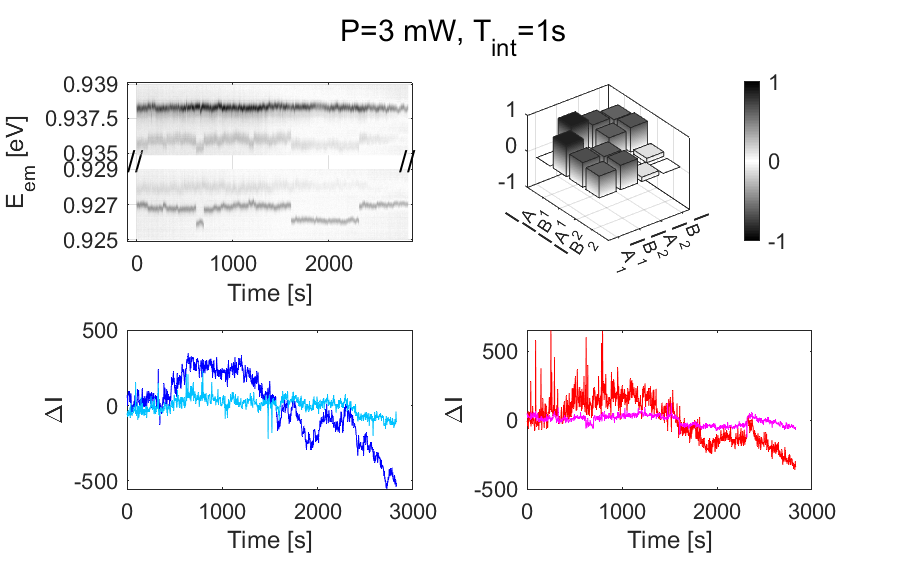}
    \includegraphics[scale=0.5]{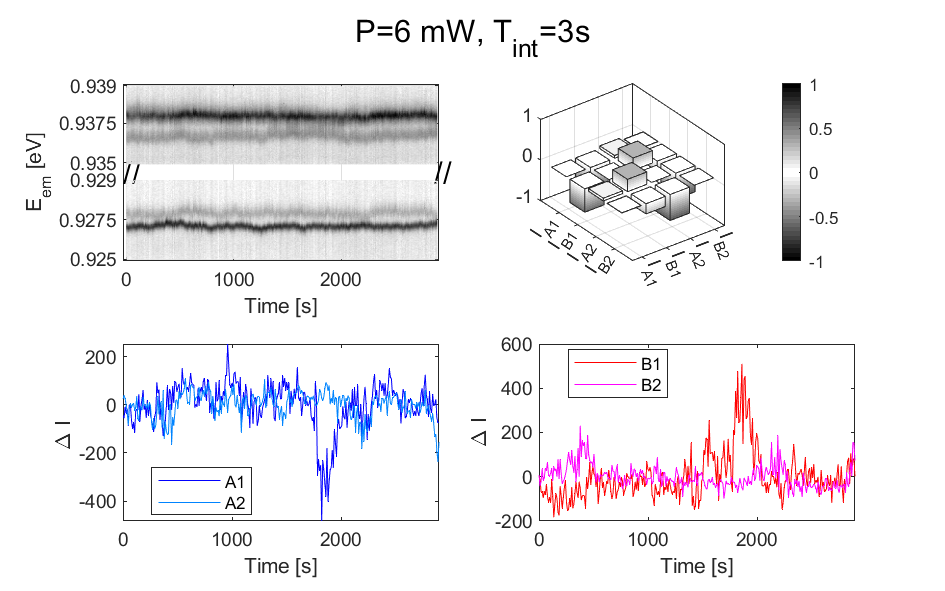}
    \includegraphics[scale=0.5]{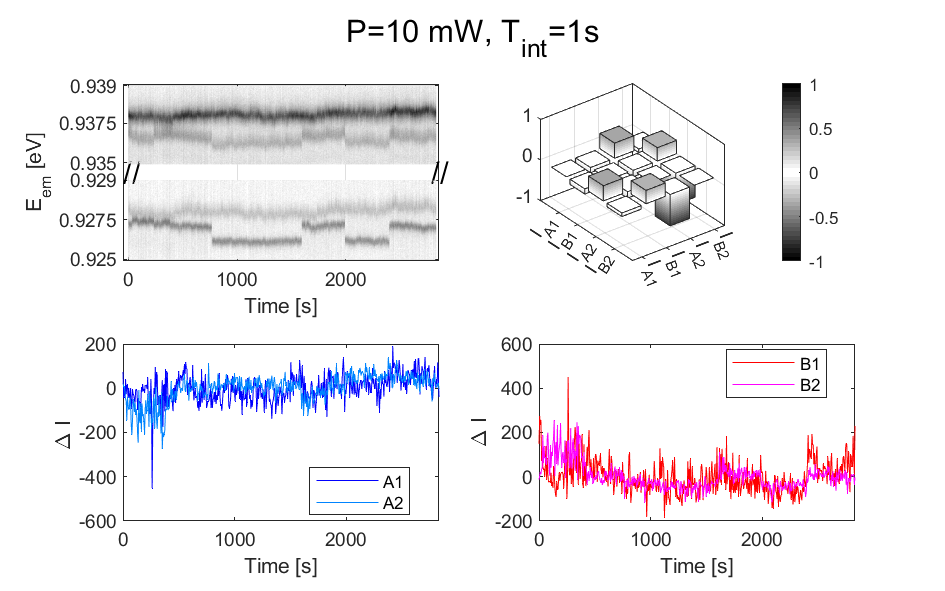}
    \includegraphics[scale=0.5]{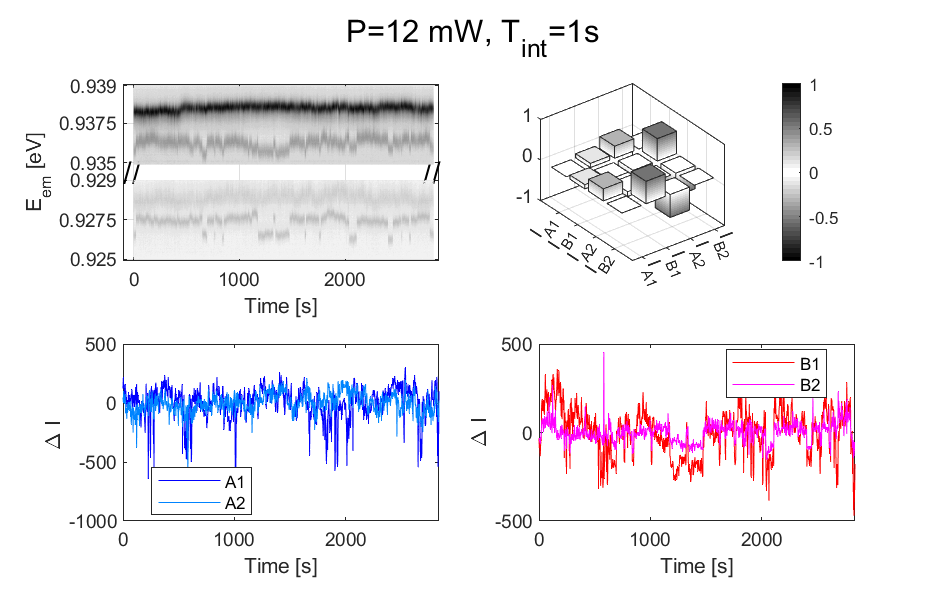}
    \includegraphics[scale=0.5]{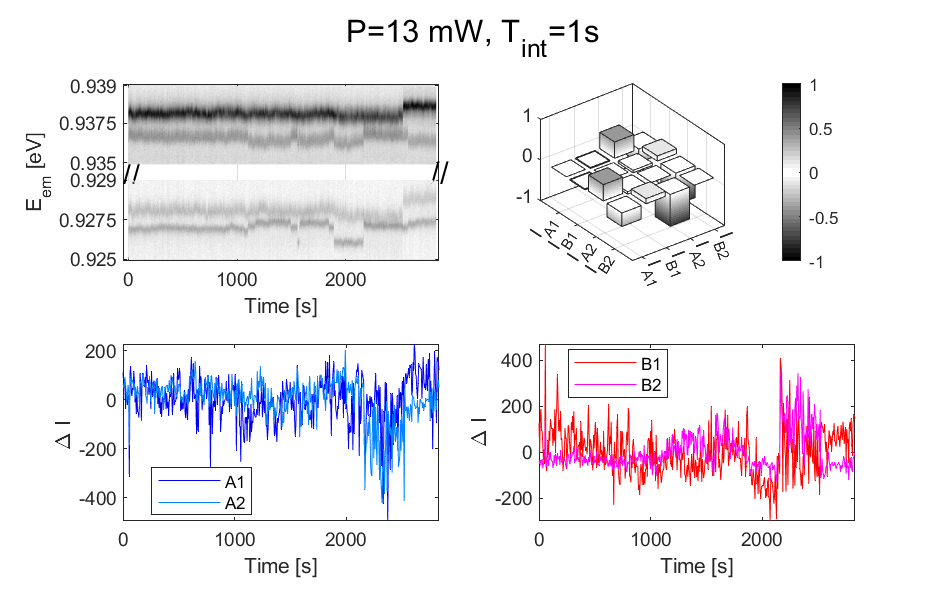}
    
    \caption{Temporal correlations between intensities for pair A and B for different excitation power. For each power, the time trace is displayed as a map in gray color on a reduced scale. Intensity deviation to the mean value $\Delta I(t)=I(t)-<I>$ is then displayed for both pair (coloured 2D plot) before computing the Pearson correlation coefficients. The data show no significant correlation between the intensities of the peaks neither within a pair nor between them.}
    \label{fig:Icorrelations}
\end{figure}

\begin{figure}
    \centering

    \includegraphics[scale=0.5]{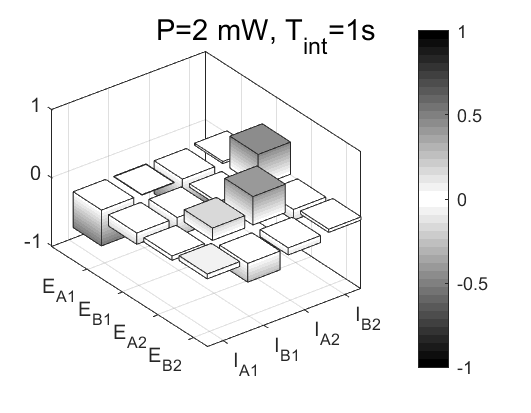}
    \includegraphics[scale=0.5]{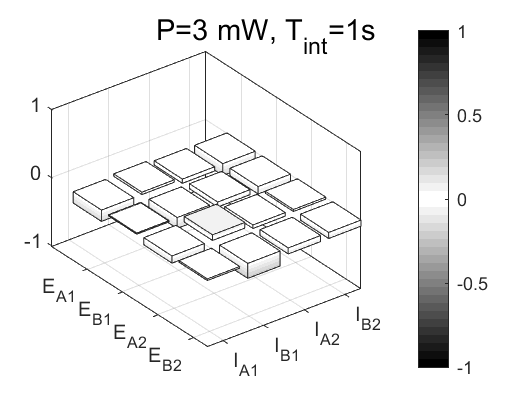}
    \includegraphics[scale=0.5]{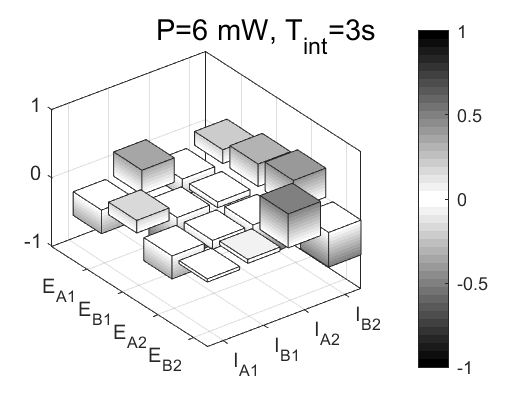}
    \includegraphics[scale=0.5]{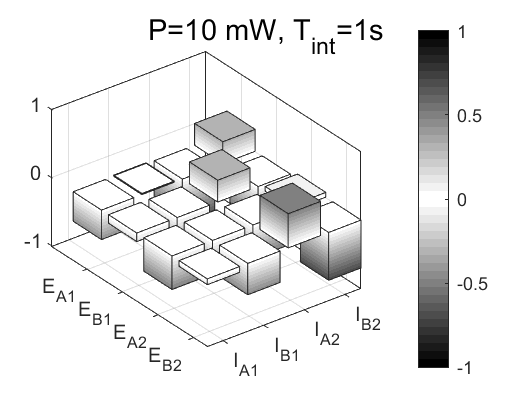}
    \includegraphics[scale=0.5]{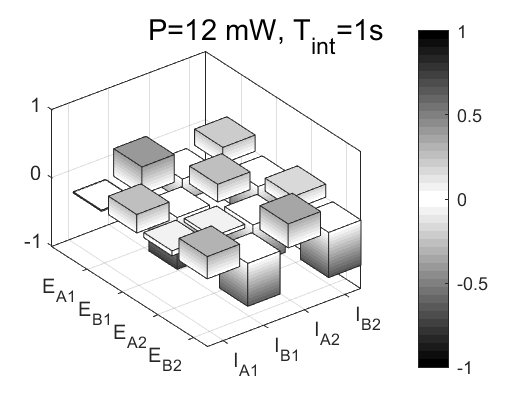}
    \includegraphics[scale=0.5]{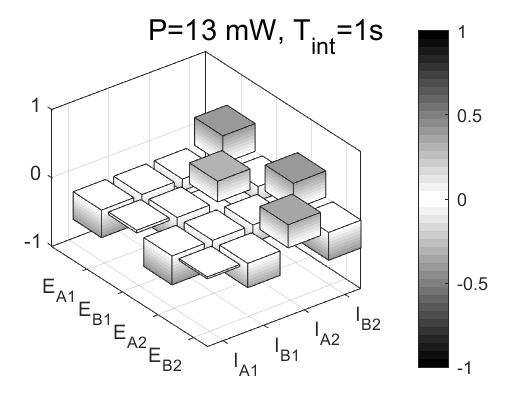}
    
    \caption{Intensity to emission energy Pearson correlations for different excitation power. The data show no significant correlations.}
    \label{fig:EIcorrelations}
\end{figure}

\clearpage

\section{Super-localization}

The super-localization technique is used to locate a nano-object with a precision much below the diffraction limit. It consists in measuring the PL intensity while scanning the sample spatially in a confocal setup. The resulting map is the convolution of the emitter geometry with the point spread function (PSF) of the optical setup. For a point-like emitter (such as a nano-emitter) the PL map boils down to the PSF. This latter can be modeled by a 2D Gaussian function, which in the appropriate (eigen) frame, reads:
\begin{equation}
    S(x,y)= A\exp(-\tfrac{(x-x_0)^2}{2\sigma_x^2})\exp(-\tfrac{(y-y_0)^2}{2\sigma_y^2}) + B,
\end{equation}
where $A$ is the amplitude, $x_0$ and $y_0$ are the spatial coordinates of the emitter, $\sigma_{x,y}$ is the width of the Gaussian function in the $x$ and $y$ directions respectively and $B$ is a global  offset. 

In our experimental setup we coupled the emitter PL into a single mode fiber. The fiber core acts as a $10 \mu m$ pinhole ensuring spatial selectivity. The emitters are excited from the backside of the substrate with a cw 800nm laser and the PL is collected through an aspheric lens (Thorlabs C330-TND-C, NA=0.7). The core of the fiber is thus conjugated to the collection spot on the sample which can be scanned using a fast steering mirror in the 4f configuration. 

\begin{figure}
    \centering
    \includegraphics[scale=0.7]{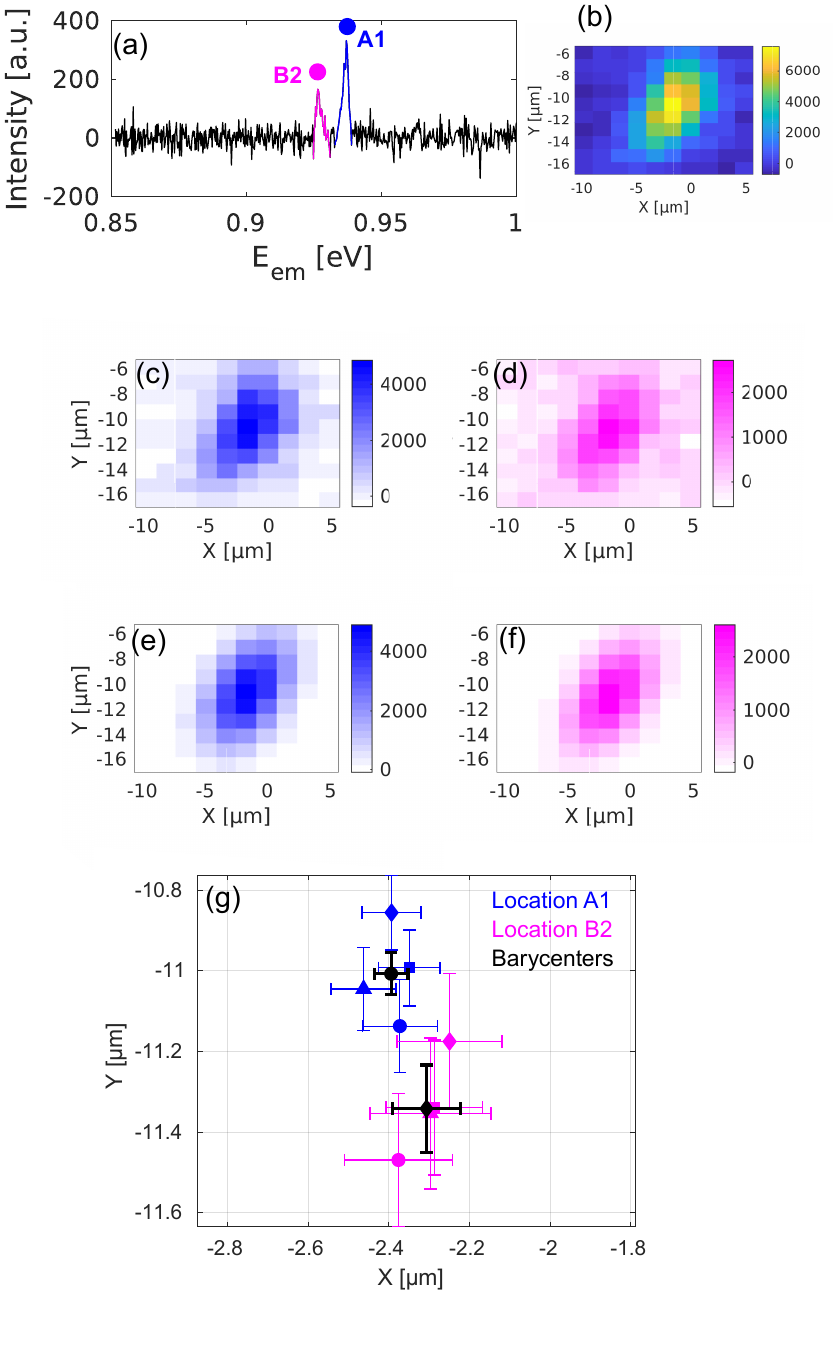}
    \caption{Superlocalization of emitter A1 and B2 using a single mode fiber as a pinhole. (a) Typical spectrum recorded with a 150 lines/mm grating. (c-d) Spectrally resolved PL maps of  emitters A1 and B2. (e,f) 2D Gaussian fits of the maps in (c) and (d). (g) Centers of the Gaussian fits, representing the super-localisation of the emitters A1 and B2 for 4 acquisitions realized in a  row. The barycenter is figured in black. Error bars resulting from the Gaussian fits.}
    \label{fig:superloc}
\end{figure}

To get the best of the superlocalization technique, the measurement should last less than the typical mechanical drift. Due to the low signal level, we chose to work with a 150 lines/mm grating to reduce the integration time down to 7s per spectra, for a typical mapping time of 15 minutes. Thus, only the two brightest lines, A1 and B2, could be resolved (Figure~\ref{fig:superloc}(a)). 

The spectrally resolved map of emitter A1 (resp. B2) is displayed in Figure~\ref{fig:superloc}.(c) (resp. \ref{fig:superloc}(d)). These maps were fitted by a 2D Gaussian function as shown in figure~\ref{fig:superloc}.(e-f). The center of the Gaussian is reported in map (g). The measurement was repeated 4 times successively to assess the mechanical drift. Each measurement is identified by a different symbol among which the color circle (resp. diamond) symbol stands for the result of the first (resp. last) measurement. We can observe for both emitters a typical drift of 200 nm in the Y direction between the first and last measurements. The barycentres are represented in black. We end up with typical distance of 300 $\pm$ 150 nm between the emitters, which allow us to unambiguously confirm that the two pairs A and B are well separated and can be considered as independent.

\section{Couple B: possible 3D electrostatic configuration}
\begin{figure}[!ht]
    \centering
    \includegraphics[width=15cm]{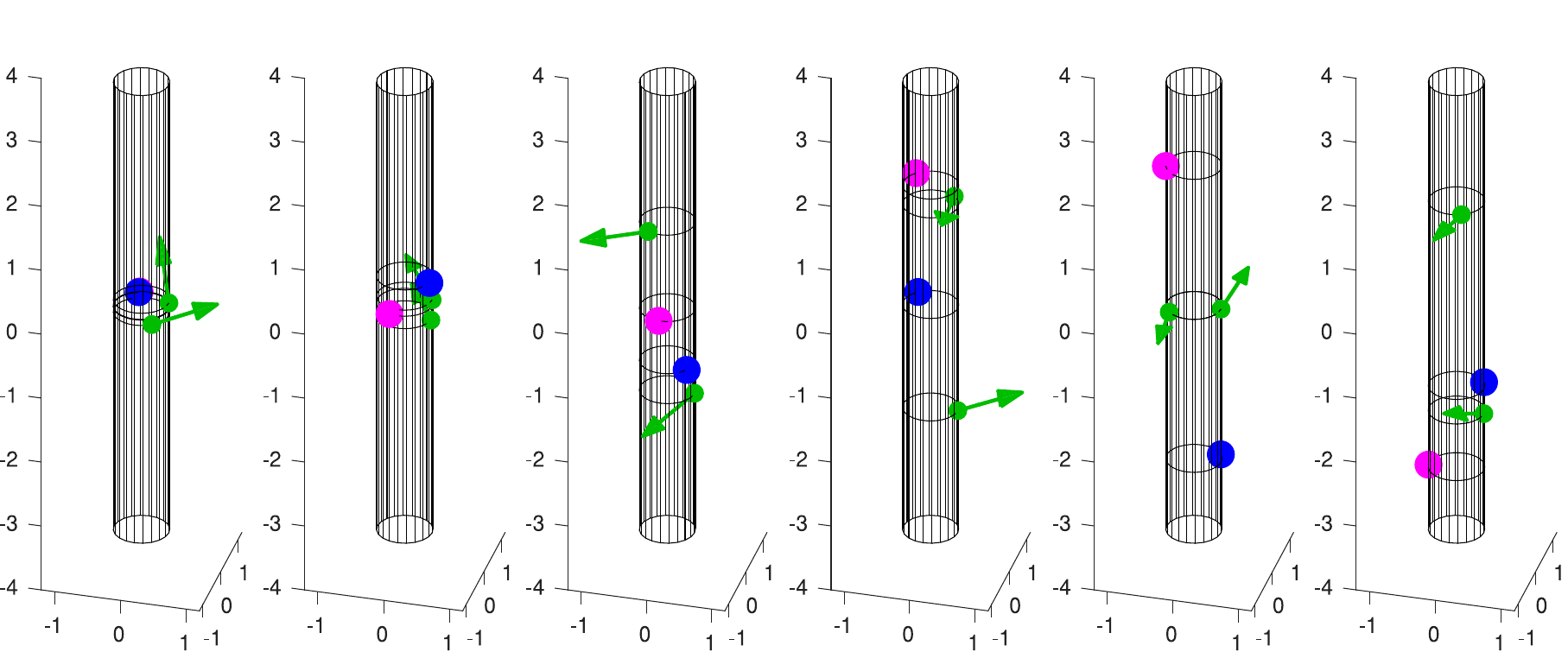}
    \caption{Six examples of possible 3D configurations compatible with the spectral diffusion observed in pair B. The green arrow indicates the permanent dipole. The blue (resp. magenta) dot represents the position of the trap corresponding to zone Z2 (resp. zone Z3) of Figure 3 of the main text.}
    \label{fig:3Dpart1}
\end{figure}

In Figure~\ref{fig:3Dpart1}, some possible 3D configurations of the dipoles associated with pair B are presented, that are compatible with the experimental data presented in Figure 3 of the main text. The average slope of the central lobe (Z1) (corresponding to trapping and detrapping of remote charges) sets the ratio of the projections of the two dipoles along the nanotube axis. The lobes Z2 and Z3 correspond to charges in the vicinity of the dipoles, for which the 3D geometry matters. We search numerically for such a 3D geometry implying the orientation and position of the two dipoles and the position of the two charge traps responsible for the lobes Z2 and Z3. A range of possible configurations are found among which we show here six examples. 

\begin{figure}[!ht]
    \centering
    \includegraphics[width=15cm]{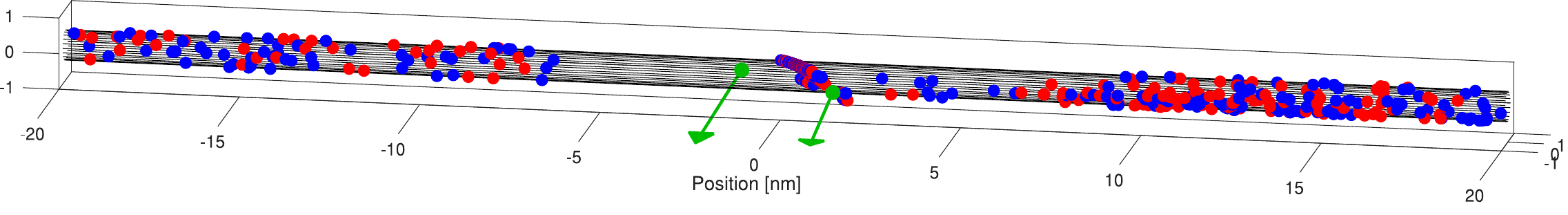}
    \caption{Each blue (trapping) or red (detrapping) dots correspond to an experimental point of fig3c of the main text for the third configuration in fig.~\ref{fig:3Dpart1}.}
    \label{fig:3Dpart2}
\end{figure}

To further confirm the adequacy of such a configuration with the whole set of experimental data, we try to reproduce directly all the experimental points observed in Figure 3 of the main paper. To do so, for each of the configurations found in fig \ref{fig:3Dpart1}, we search numerically for the positions of the trapping or detrapping events responsible for the main lobe Z1, using trap positions  as close as possible to the nanotube surface (as an example see Figure \ref{fig:3Dpart2}, corresponding to the third dipole configuration displayed in Figure \ref{fig:3Dpart1}).

\section{Autocorrelation function for a fast switching emitter}

In this section we compute the autocorrelation function of a two-level system emitter that switches between two configurations, resulting in two distinct emission lines when observed with a long integration time. Since we observe several time scales from the experimental autocorrelation function, we will first show that the autocorrelation function can be factorised. We will then explicitly calculate the part of $g^{(2)}$ corresponding to a fast switching emitter.

\subsection{Factorization of the $g^{(2)}$ function}
Here we will show that if the different time scales involved in the $g^{(2)}$ function are very different, the $g^{(2)}$ function can be factorised.  This procedure has already been used by Nutz and coworker \cite{Nutz2019} , who showed that the behaviour of an anti-bunching and a bunching originating from a dark state can be well described by treating them separately.

Following the work of Fishman and co-workers \cite{Fishman2023}, by introducing $\vec{P}$: a vector of state occupancy probabilities, and $G$ the transition rate matrix, the system of rate equations is given by $\frac{d}{dt}\vec{P} = G \vec{P}$. The formal solution is given by:
\begin{equation}
\vec{P}(t) = (\vec{w}_0\cdot \vec{P}_0) \vec{v}_0 +  \sum_{i=1}^{n-1} (\vec{w}_i\cdot \vec{P}_0) e^{\lambda_i t} \vec{v}_i,
\end{equation}
where $\vec{P}_0$ is the initial state, $\vec{v}_i$ are the eigenvectors of $G$ with the associated eigenvalue $\lambda$: $G\vec{v}_i = \lambda_i \vec{v}_i$. The vector $\vec{v}_0$ is the eigenvector with null eigenvalue, corresponding to the stationary solution.  The vector $\vec{w}_j$ corresponds to the dual basis; $\vec{w}_j\cdot \vec{v}_i = \delta_{i,j}$. 
The autocorrelation function is given by the conditional probability of being in the excited state $e$ at time $\tau$ under the condition of being in the ground state $g$ at $\tau=0$ and normalised by the steady state population of $e$:
\begin{equation}
g^{(2)}(\tau) = 1 +  \sum_{i=1}^{n-1} \frac{(\vec{w}_i\cdot \vec{g})(\vec{e}\cdot \vec{v}_i)}{(\vec{w}_0\cdot \vec{g})(\vec{e}\cdot \vec{v}_0) } e^{-\tau/\tau_i} .
\end{equation}
Since the real part of all eigenvalues is negative or zero  \cite{Fishman2023}, we set $\tau_i = -1/Re(\lambda_i)$ . We also assume that all characteristic times are very distinct: $\tau_1 \ll \tau_2 \ll ..\ll \tau_{n-1}$ .  By using $e^{-t/\tau_i}e^{-t/\tau_{n-1}} \simeq e^{-t/\tau_i}$, the autocorrelation function can be approximated by:
\begin{equation}
g^{(2)}(\tau) \simeq \Bigg( 1 +  \sum_{i=1}^{n-2} \frac{(\vec{w}_i\cdot \vec{g})(\vec{e}\cdot \vec{v}_i)}{(\vec{w}_0\cdot \vec{g})(\vec{e}\cdot \vec{v}_0) + (\vec{w}_{n-1}\cdot \vec{g})(\vec{e}\cdot \vec{v}_{n-1}) } e^{-\tau/\tau_i}  \Bigg) \Bigg(1 + \frac{(\vec{w}_{n-1}\cdot \vec{g})(\vec{e}\cdot \vec{v}_{n-1})}{(\vec{w}_0\cdot \vec{g})(\vec{e}\cdot \vec{v}_0) } e^{-\tau/\tau_{n-1}} \Bigg).
\end{equation}
Defining a new steady state vector: 
\begin{eqnarray}
 \vec{v}_0' &\equiv & \frac{1}{\sqrt{ (\vec{w}_0 \cdot \vec{g})(\vec{v}_0 \cdot \vec{g})+ (\vec{w}_{n-1} \cdot \vec{g})(\vec{v}_{n-1} \cdot \vec{g})}} \Big( (\vec{w}_0 \cdot \vec{g}) \vec{v}_0 + (\vec{w}_{n-1} \cdot \vec{g}) \vec{v}_{n-1} \Big) , \\
\vec{w}_0' &\equiv& \frac{1}{\sqrt{ (\vec{w}_0 \cdot \vec{g})(\vec{v}_0 \cdot \vec{g})+ (\vec{w}_{n-1} \cdot \vec{g})(\vec{v}_{n-1} \cdot \vec{g})}} \Big( (\vec{v}_0 \cdot \vec{g}) \vec{w}_0 + (\vec{v}_{n-1} \cdot \vec{g}) \vec{w}_{n-1} \Big) ,  
\end{eqnarray}
leads to :
\begin{equation}
(\vec{w}_0\cdot \vec{g})(\vec{e}\cdot \vec{v}_0) + (\vec{w}_{n-1}\cdot \vec{g})(\vec{e}\cdot \vec{v}_{n-1}) = (\vec{w}_0'\cdot \vec{g})(\vec{e}\cdot \vec{v}_0').
\end{equation}
It can be verified that the basis $\vec{v}_0'$ , $\vec{v}_1$, $\vec{v}_2$ ... $\vec{v}_{n-2}$  and its dual $\vec{w}_0'$ , $\vec{w}_1$, $\vec{w}_2$ ... $\vec{w}_{n-2}$ form an orthonormal basis of dimension $n-2$.  We end up with the autocorrelation function given by:
\begin{equation}
g^{(2)}(\tau) \simeq \Bigg( 1 +  \sum_{i=1}^{n-2} \frac{(\vec{w}_i\cdot \vec{g})(\vec{e}\cdot \vec{v}_i)}{(\vec{w}_0'\cdot \vec{g})(\vec{e}\cdot \vec{v}_0') } e^{-\tau/\tau_i}  \Bigg) \Bigg(1 + \frac{(\vec{w}_{n-1}\cdot \vec{g})(\vec{e}\cdot \vec{v}_{n-1})}{(\vec{w}_0\cdot \vec{g})(\vec{e}\cdot \vec{v}_0) } e^{-\tau/\tau_{n-1}} \Bigg).
\end{equation}
The left part corresponds to the $g^{(2)}$ function of an effective system of dimension $n-1$, and the right part corresponds to the $g^{(2)}$ function of a two-level system.  We note that the state $g$ and $e$ belongs to the full system defined by the state $\vec{v}_n$ where $n$ ranges from 0 to $n-1$. Only the projection on each subsystem is required, and thus we really end up with a factorisation of the $g^{(2)}$ function.

By reiterating this procedure, the autocorrelation function can be factorised into a product of  $g^{(2)}$ of simple system consisting of two effective levels associated with a characteristic time $\tau_i$:
\begin{equation}
g^{(2)}(\tau) \simeq \prod_{i=1}^{n-1} g^{(2)}_i(\tau) .
\end{equation}

Assuming a fast switching of the emitter between two states, even in the presence of other levels (dark states, etc.), one can focus on the sub $g^{(2)}$ corresponding to this switching time-scale.

\subsection{Part of the autocorrelation function that corresponds to fast switching.}

\begin{figure}[ht!]
    \centering
    \includegraphics[scale=0.5]{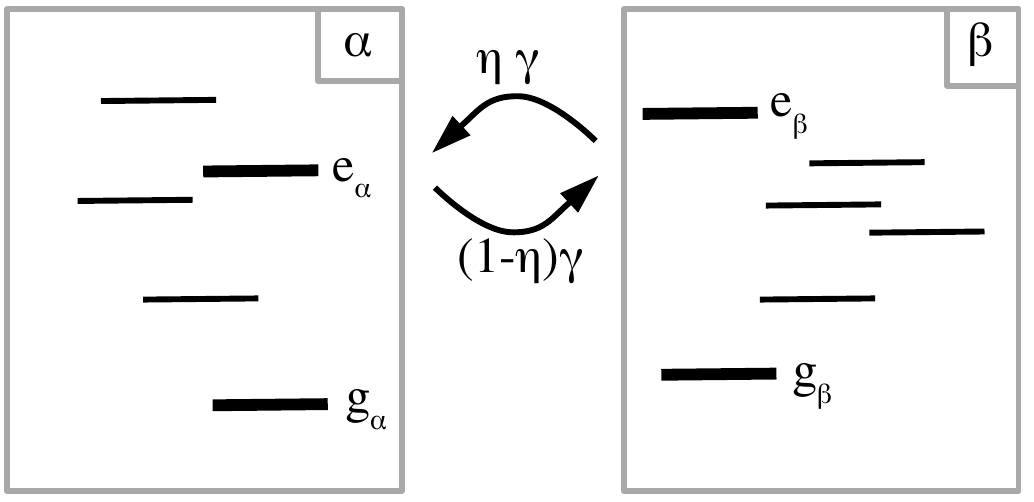}
    \caption{  Electronic configuration for a hypothetical fast switching between two configurations $\alpha$ and $\beta$. The optical transition comes from the transitions $e_\alpha \to g_\alpha$ and $e_\beta \to g_\beta$.  }
    \label{fig:switch}
\end{figure}

In the subspace correponding to the fast switching between the configuration $\alpha$ and configuration $\beta$, cf figure \ref{fig:switch}, the transition matrix reads:
\begin{equation}
G=\gamma \left( \begin{array}{cc}
-(1-\eta) & \eta \\
1-\eta & -\eta 
\end{array}\right),
\end{equation}
the eigenvectors are given by $\vec{v}_0 = (\eta ,1-\eta)$ ; $\vec{v}_1 = (1,-1)$ ; $\vec{w}_0 = (1,1)$ ; $\vec{w}_1=(1-\eta,-\eta)$. The eigenvector $\vec{v}_0$ corresponds to the stationary solution and $\vec{v}_1$ to the eigenvalue $-\gamma$ . 
The part of the intensity autocorrelation of the transition $e_\alpha \to g_\alpha$  related to the fast switching  corresponding, is given by:
\begin{equation}
g^{(2)}_{\alpha}(\tau) = 1 +\frac{(\vec{w}_{1}\cdot \vec{g}_\alpha)(\vec{e}_\alpha\cdot \vec{v}_{1})}{(\vec{w}_0\cdot \vec{g}_\alpha)(\vec{e}_\alpha\cdot \vec{v}_0) } e^{-\gamma\tau}.
\end{equation}
The expression of $g^{(2)}_{\beta}$ has the same expression as $g^{(2)}_{\alpha} $ by replacing $\alpha$ with $\beta$ .  Noticing that $e_\alpha$ and $g_\alpha$ belong to the subspace $\alpha$ ; and $e_\beta$ and $g_\beta$  to the subspace $\beta$ ; we finally obtain:  
\begin{eqnarray}
g^{(2)}_{\alpha}(\tau) = 1 + \frac{1-\eta}{\eta} e^{-\gamma |\tau|}, \\
g^{(2)}_{\beta}(\tau) = 1 + \frac{\eta}{1-\eta} e^{-\gamma |\tau|} .
\end{eqnarray}

A fast switching event is thus characterized by a step (in semi-logx scale) in the auto-correlation of both lines, at the same characteristic time $1/\gamma$. Moreover the product of the amplitudes of this step of each line is equal to one (for a normalized $g^{(2)}$ function).

Note that the presence of oscillations in the autocorrelations for times exceeding 0.1 ms are signatures of mechanical vibrations in our setup due to the high sensitivity of the Fabry-Perot cavity used for spectral filtering. 


\section{1D vs 3D emitter pair generation probability}
For an emitter numeric concentration $\rho$, the probability to find $k$ emitters within an elementary volume $v$ reads: $P_k(v) = e^{-v\rho} (v\rho)^k/k!$. The manipulation of a quantum coupled pair requires that two emitters are in close vicinity (within $dv~$ a few nm$^3$) and are the only pair within a volume $V_\lambda$ of the order of $\lambda^3$, where $\lambda$ is the transition wavelength used to manipulate the system. This configuration has a probability $P = \frac{V_\lambda}{dv} P_2(dv) P_0(V_\lambda-dv)$. This probability is maximum for a concentration $\rho=2/V_\lambda$ and amounts to $P_{max} = \frac{dv}{V_\lambda} 2 e^{-2}$.
In a 3D system, with $dv=\delta^3$, this reads: $P_{max}\simeq0.27 (\delta/\lambda)^3$, whereas for a 1D system, one gets $P_{max}\simeq0.27 \delta/\lambda$. Hence, for typical configurations where $\delta/\lambda\simeq 0.001$, the 1D configuration is $10^6$ times more favorable than its 3D counterpart, showing the strong interest of using a nanotube as a 1D backbone to anchor the OCCs.

\section{Other example of coupled pair.}
\begin{figure}[ht!]
    \centering
    \includegraphics[scale=0.5]{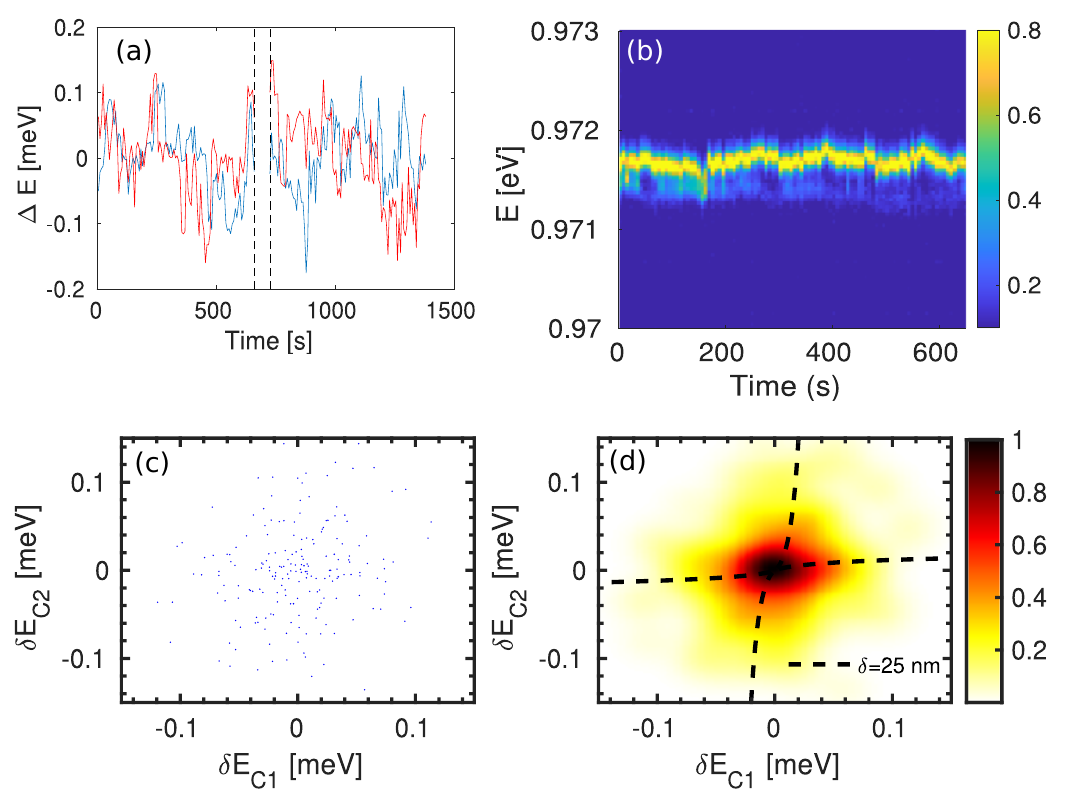}
    \caption{  (a) Time trace of energy shifts (b) Time trace of the PL spectrum (c) Spectral shifts for the emitters of pair $C$ (d)  Density plot generated from panel (c). The black dashed line corresponds to the projection of the permanent dipole:  $\vec{p}_{C1}\cdot\vec{e}_z= |p_0|$ and  $\vec{p}_{C2}\cdot\vec{e}_z= 0.75|p_0|$. The dipole separation is $\delta=25 \ nm$.}
    \label{fig:otherexample}
\end{figure}

In figure \ref{fig:otherexample}, we show another pair of OCCs on a single nanotube (we have named this pair "C"), similar to those presented in the main text. From the electrostatic correlation, we found that the distance between the two OCCs is in the order of $25 \ nm$.

\section{Instrumental response function}

To measure the instrumental response function, we use a homemade super-continuum laser source. A non-linear photonic fiber (NKT Photonics NL-PM-750) is pumped with a femtosecond pulsed laser (Coherent Vitesse) to generate a super-continuum in the range 500 nm to 1100 nm. The output light is filtered out using hard coated 10 nm bandpass filter and coupled to a single-mode fiber plugged to our single photon detector. Simulations show that for a commercial fiber with a typical length of 10 m, we expect femtosecond pulses to be broadened up to few picoseconds. Since the pulse width is an order of magnitude smaller than the jitter time, we consider them as delta functions in the time domain, so that :

\begin{equation}
    R_{mes}(t) = (R_{SSPD}*S_{pulse})(t) \simeq (R_{SSPD}*\delta)(t) = R_{SSPD}(t) 
\end{equation}

\begin{figure}
    \centering
    \includegraphics{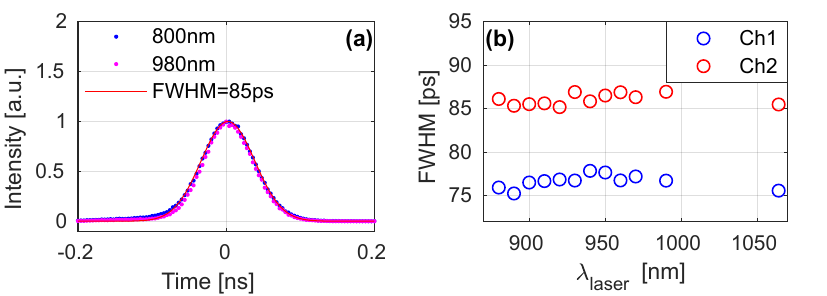}
    \caption{Single photon detector time response measured for two different wavelengths with gaussian fit (a). Extracted FWHM from the fit with respect to the laser wavelength for both channels}
    \label{fig:SI_IRF}
\end{figure}

We first characterized our detector by measuring the $R_{mes}$ with respect to the excitation wavelengths in the range 800 nm to 1100 nm, as shown on figure \ref{fig:SI_IRF}.(b). The 800 nm point is made by coupling the femtosecond laser directly into a single mode fiber and plugged to the single photon detector to ensure that the pulse broadening induced by the non-linear optical fiber is negligible. The full width at half maximum (FWHM) is constant for each channel over the wavelength. Although jitter is not well understood, it is intrinsic to each detector and tends to be identical over the designed range of bias current and wavelength. Based on this understanding, we keep the jitter constant over the full range of our detectors. 

\input{SI.bbl}

%% file: MAIN.bbl
\providecommand{\latin}[1]{#1}
\makeatletter
\providecommand{\doi}
  {\begingroup\let\do\@makeother\dospecials
  \catcode`\{=1 \catcode`\}=2 \doi@aux}
\providecommand{\doi@aux}[1]{\endgroup\texttt{#1}}
\makeatother
\providecommand*\mcitethebibliography{\thebibliography}
\csname @ifundefined\endcsname{endmcitethebibliography}
  {\let\endmcitethebibliography\endthebibliography}{}

%% file: SI.bbl
%